\documentclass[11pt]{article}
\usepackage{jb}

\renewcommand{\labelitemi}{$\bullet$}

\usepackage{amssymb}

\newcommand{\bi}{\begin{itemize}}
\newcommand{\ei}{\end{itemize}}
\newcommand{\be}{\begin{enumerate}}
\newcommand{\ee}{\end{enumerate}}
\newcommand{\bd}{\begin{description}}
\newcommand{\ed}{\end{description}}

\newcommand{\mca}[2]{\multicolumn{#1}{@{}l@{}}{#2}}

\renewcommand{\.}[1]{\!#1\!}

\newlength{\beweisskip}
\newlength{\blindlgth}
\newlength{\addblindlgth}

\newcommand{\ra}{\rightarrow}

\newcommand{\lra}{\leftrightarrow}

\newcommand{\COMMENT}[1]{}

\renewcommand{\phi}{\varphi}

\renewcommand{\leq}[0]{\leqslant}
\renewcommand{\geq}[0]{\geqslant}
\newsavebox{\ninbox}
\savebox{\ninbox}{{\scriptsize $\not\in$}}

\newsavebox{\sinbox}
\savebox{\sinbox}{{$\scriptscriptstyle\in$}}
\newsavebox{\inbox}
\savebox{\inbox}{{\scriptsize $\in$}}

\newsavebox{\upinbox}
\savebox{\upinbox}{%
	$\cup%
	\hspace*{-0.370em}%
	\rule{0.02cm}{1.25ex}%
	\hspace*{0.370em}$}

\renewcommand{\in}
	{\mathrel{\mathchoice{\raisebox{0.5mm}{\usebox{\inbox}}}
			     {\raisebox{0.5mm}{\usebox{\inbox}}}
			     {\raisebox{0.0mm}{\usebox{\sinbox}}}
			     {\raisebox{0.0mm}{\usebox{\sinbox}}}}}

\newsavebox{\greybox}
\newlength{\greygrain}
\newcounter{poster}
\renewcommand{\:}[4]{%
        {%
        \renewcommand{\:}[4]{%
                {%
                \renewcommand{\:}[4]{error\error}%
                \renewcommand{\j}{{##2}}%
                {##4}%
                ##1...##1%
                \renewcommand{\j}{{##3}}%
                {##4}%
                }%
        }%
        \renewcommand{\i}{{#2}}%
        {#4}%
        #1\ldots#1%
        \renewcommand{\i}{{#3}}%
        {#4}%
        }%
}

\newcommand{\pfundef}[3]{%
\Rnewcommand{#1}[1]{#2##1#3}%
}

\pfundef{\set}{\{}{\}}                  
\pfundef{\tpl}{\langle}{\rangle}        
\pfundef{\bag}{[\![}{]\!]}              
\pfundef{\subst}{\{}{\}}                

\newcommand{\SECT}[1]{\section{#1}}		
\newcommand{\SSCT}[1]{\paragraph{{#1.}}}	
\newcommand{\COMM}[1]{}				
\newcommand{\EMPH}[1]{{\em #1}}			
\newcommand{\FORM}[1]{$#1$}			
\newcommand{\QUOT}[1]{``#1''}			
\newcommand{\CITE}[1]{[#1]}			
\newcommand{\BI}{\begin{itemize}}		
\newcommand{\EI}{\end{itemize}}			
\newcommand{\ITEM}{\item}			
\newcommand{\BD}{\begin{description}}		
\newcommand{\ED}{\end{description}}		
\newcommand{\DITM}[1]{\item[{\rm [#1]}]}	

\newcommand{\eqd}[1]{eqd#1}
\newcommand{\eqr}[1]{eqr#1}

\begin{document}

\begin{center}
{\Large\bf Experiences in Developing Time-Critical Systems	\\
--
The Case Study ``Production Cell''

}
\end{center}
\vspace{1.0cm}

Jochen Burghardt
\hfill
1994
\vspace{1cm}


\SECT{Aim}
\COMM{- what was the aim of the work (a running controller? a verified
	controller? a formal description of...?)}

Our aim is to write a formal specification of the production cell that
is as close as possible to the informal requirements description and to
show how a verified TTL-like circuitery can be constructed from this,
using deductive program synthesis. The main emphasis lies on the formal
requirements specification which also covers mechanical aspects and thus
allows to reason not only about software issues but also about issues of
mechanical engineering.

Besides an approach confined to first order predicate logic with
explicit, continuous time, an attempt is presented to employ application
specific user defined logical operators to get a more concise
specification as well as proof.

\SECT{Deductive program synthesis}
\COMM{Kurze Darstellung der angewandten Methode/Sprache}

The deductive program synthesis approach due to Manna and Waldinger
\CITE{Manna80}
is a method for program development in the small. No methodological
support e.g. for decomposition into modules is provided, instead, it
concentrates on deriving one algorithm from a given specification and
some given axioms of background knowledge.

Axioms and specifications are given as first order predicate logic
formulas. One tries to prove the specification formula, thereby
simultaneously constructing a correct functional program from the answer
substitutions arising from unification. The proof rules include
resolution for formulas in non-clausal form and some generalisations of
e-resolution and paramodulation described in
\CITE{Manna86}.

A program is purely functional and represented as a term built up from
function symbols including a ternary
\FORM{if \cdot then \cdot else \cdot fi},
recursive programs arise from induction in the proof.

Figure 
\ref{@F1@}
shows as a very easy example the synthesis of a program
\FORM{x}
satisfying the specification
\FORM{p(x)}.
Formulas 1 and 2 provide the assumed background knowledge, formula 3
states the proof goal. The resolution proof in step 4. - 6. actually
proves the formula
\FORM{\exists x \; p(x)}.
The
\QUOT{output}
column of formula 6 contains the synthesized program.

	\begin{figure}
	\COMM{10.5cm x 5.5cm File f1.ps}
	{\normalsize%
%
%
%
%
%
%
%
%

\begin{center}


\begin{tabular}{@{}|r||l|l|l||l|@{}}
\hline
& Assertions & Goals & Output &	\\
\hline
\hline
1. & $q(x) \ra p(f(x))$ &&& Axiom	\\
\hline
2. & $q(c) \vee q(d)$ &&& Axiom	\\
\hline
3. && $p(x)$ & $x$ & Specification	\\
\hline
\hline
4. && $q(x_1)$ & $f(x_1)$ & 3 res 1	\\
\hline
5. && $\neg q(d)$ & $f(c)$ & 4 res 2	\\
\hline
6. && $true$ &
	\begin{tabular}[t]{@{}l@{$\;$}l@{}}
	if & $q(d)$	\\[-0.05cm]
	then & $f(d)$	\\[-0.05cm]
	else & $f(c)$	\\ [-0.05cm]
	fi	\\[-0.05cm]
	\end{tabular}
	& 5 res 4	\\
\hline
\end{tabular}

\end{center}

%
%
%
%
%
%
%
%
%
%
%
%
%

\par}
	\vspace{0.5cm}

	\caption{Example synthesis proof}
	\label{@F1@}
	\end{figure}

\SECT{Modelling the production cell}
\label{Modelling the production cell}
\COMM{Wie wurde die Methode/Sprache angewandt? - Darstellung der
	grundsaetzlichen Vorgehensweise}

It is well known that the transition from an informal requirements
description to a formal specification is the most critical step wrt.
correctness within the formal scenario, since the formal specification
can of course not be mathematically verified against the informal
description. We tried to adopt an approach to defuse this problem as far
as possible: to create a formal language level in which the informal
description 
can be expressed almost
\QUOT{1:1}
and thus be easily validated. The specification obtained this way is a
requirements specification, not a design specification; due to its high
degree of implicitness it does not admit rapid prototyping, nor an
immediate stepwise refinement into executable code.

First, a suitable terminology has been fixed, consisting of predicate
and function symbols together with their informal explanations. Figure
\ref{@F2@}
shows some example explanations. Time has been modelled by explicit
parameters in order to cope with the restriction to first order
predicate logic, and to be able to talk about deadlines explicitly.
Space is modelled by three-dimensional cartesian coordinate vectors,
transformation into, resp. from, polar coordinates are axiomatized as
far as needed. The desired 
\QUOT{program}
will consist in an asynchronous circuitery built up from TTL-like
components, modelled as time-dependent functions. Switching times are
ignored within this setting. No explicit feedbacks are allowed in the
circuitery, since this would amount to deal with infinite terms which is
not supported by the proof tool. Instead, circuitery feedbacks are
hidden in circuits like flip flops.

	\begin{figure}
	\COMM{14.5cm x 7.0cm File f2.ps}
	{\normalsize%
%
%
%
%
%
%
%
%

\begin{center}


\begin{tabular}{@{}l@{$\;$}c@{$\;$}l@{}}
\mca{2}{{\bf Prediactes:}}	\\
$robot(r,x)$ & $\lra$ & $r$ is a two armed robot placed 
	at coordinates $x$	\\
$extends_i(r,t)$ & $\lra$ & at time $t$, the robot $r$ is extending its
	$i^{\mbox{\scriptsize th}}$ arm	\\[0.1cm]
\mca{2}{{\bf Functions:}}	\\
$pos_i(r,t)$ & $=$ & coordinates of the electromagnet of robot $r$'s
	$i^{\mbox{\scriptsize th}}$ arm at time $t$	\\
$dist\_xy(x,x_1)$ & $=$ & distance of the $xy$ projections of 
	coordinates $x$ and $x_1$ \\
$r(\vec{c},\vec{s})$ & $=$ & a two armed robot with control inputs
	$\vec{c}$ and sensor outputs $\vec{s}$	\\
$val(c,t)$ & $=$ & value of the time-dependent function $c$ at time 
	$t$ \\
$trigger(c,v)$ & $=$ & output of a Schmitt trigger circuit with input
	$c$ and threshold $v$	\\
\mca{2}{{\bf Constants:}}	\\
$d_3$ & $=$ & coordinates of the elevating rotary table 
	(turning center) \\
$d_4$ & $=$ & coordinates of the robot (turning center)	\\
$maxlg_i$ & $=$ & maximum length the $i^{\mbox{\scriptsize th}}$ arm 
	of a robot can extend to	\\
\end{tabular}

\end{center}

%
%
%
%
%
%
%
%
%
%
%
%
%

\par}
	\vspace{0.5cm}

	\caption{Informal meanings of some predicate, function, and
			constant symbols}
        \label{@F2@}
	\end{figure}

\begin{figure}
	\COMM{16.0cm x 14.2cm File f3.ps}
	{\normalsize
%
%
%
%
%
%
%
%
%

\renewcommand{\eqd}[1]{\item[\bf #1:]}


{\bf Module ``Robot'':}

\bi

\eqd{11}
If the first arm is extending long enough, it will eventually reach 
each length between its current and its maximal one. 

\begin{tabular}[t]{@{}l@{}r@{$\;\;$}l@{}}
$\forall r,x,t,d \; \exists t_2:$
	&& $robot(r,x)$	\\
	& $\wedge$
	& $dist\_xy(x,pos_1(r,t)) \.\leq d \.\leq maxlg_1$ \\
& $\ra$ &	\begin{tabular}[t]{@{}l@{}r@{$\;\;$}l@{}l@{}}
	$($ && $(\forall t_1: \;\; t \.\leq t_1 \.< t_2
		\;\ra\; extends_1(r,t_1))$ \\
	& $\ra$ & $dist\_xy(x,pos_1(r,t_2)) \.= d$ & $)$ \\
	\end{tabular}	\\
& $\wedge$ & $(\forall t_3: \; t \.\leq t_3 \.< t_2 
	\ra dist\_xy(x,pos_1(r,t_3)) \.< d)$	\\
\end{tabular}

\eqd{12}
Only if the first arm extends, its length can grow.

\begin{tabular}[t]{@{}l@{$\;\;$}r@{$\;\;$}l@{}}
$\forall r,x,t,t_2:$ 
	&& $robot(r,x)$	\\
	& $\wedge$ & $t \.\leq t_2$	\\
	& $\wedge$ &
	$dist\_xy(x,pos_1(r,t)) \.< dist\_xy(x,pos_1(r,t_2))$	\\
& $\ra$ & $\exists t_1: \;\; t \.< t_1 \.< t_2 \;\wedge\;
	extends_1(r,t_1)$	\\
\end{tabular}

\eqd{13}
Motor control and sensors
($c_1$: extend first arm, $s_1$: length of first arm)

\begin{tabular}[t]{@{}r@{$\;\;$}l@{}}
\mca{2}{$\forall c_1,c_2,c_3,c_4,c_5,c_6,c_7,c_8,s_1,s_2,s_3,x,t:$} \\
& $robot(r(c_1,c_2,c_3,c_4,c_5,c_6,c_7,c_8,s_1,s_2,s_3),x)$ \\
$\ra$ & $(extends_1(r(c_1,c_2,c_3,\ldots,c_8,s_1,s_2,s_3),t)
	\;\lra\; val(c_1,t) \.= 1)$ \\
$\wedge$ & $dist\_xy(x,pos_1(
	r(c_1,c_2,c_3,\ldots,c_8,s_1,s_2,s_3),t))  
	 \.= val(s_1,t)$ \\
\end{tabular}

\ei
\vspace{0.4cm}


{\bf Module ``Factory'':}

\bi
\eqd{21}
A two armed robot is placed at $d_4$.

$robot(r(c_1,c_2,c_3,c_4,c_5,c_6,c_7,c_8,s_1,s_2,s_3),d_4)$

\eqd{22}
The elevating rotary table is reachable by the first arm of the robot.

$dist\_xy(d_4,d_3) \leq maxlg_1$

\ei
\vspace{0.4cm}


{\bf Module ``Circuits'':}

\bi

\eqd{31}
Trigger circuit
\hfill
\begin{picture}(3.2,0.1)
\put(0.600,0.625){\makebox(0.000,0.000)[r]{$in$}}
\put(0.700,0.625){\line(4,1){0.500}}
\put(1.200,0.750){\line(4,-1){0.500}}
\put(1.700,0.625){\line(4,-1){0.500}}
\put(2.200,0.500){\line(4,1){0.500}}
\multiput(0.7,0.625)(0.1,0){22}{\line(1,0){0.05}}
\put(3.000,0.625){\makebox(0.000,0.000)[l]{$v$}}
\put(0.600,0.125){\makebox(0.000,0.000)[r]{$out$}}
\put(0.700,0.000){\line(1,0){1.000}}
\put(1.700,0.000){\line(0,1){0.250}}
\put(1.700,0.250){\line(1,0){1.000}}
\end{picture}

$\forall c,v,t: \;\; val(trigger(c,v),t) \.= 1 \;\lra\; val(c,t) \.< v$

\ei
%
%
%
%
%
%
%
%
%
%
%
%
%

\par}
	\vspace{0.5cm}

	\caption{Some axioms from the specification}
	\label{@F3@}
	\end{figure}
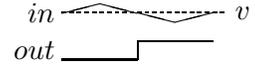

Then, a collection of obvious facts about the behaviour of the machines
could be formalised. See figure 
\ref{@F3@}
for some examples; the full specification and the synthesis proof are
contained in 
\CITE{Burghardt94}.

The formal specification consists of four parts:
\BI
\ITEM the description of behaviour required from each machine,
\ITEM the description of behaviour required from each control circuit,
\ITEM background facts from geometry, arithmetics and physics, and
\ITEM the actual specification of the production cell's goal.
\EI
The specification has the property of locality in the sense that in
order to validate a certain axiom it is only necessary to check this
single axiom against its informal description, using the terminology
description.

The specification has been modularized in the obvious way, having for
each machine type one module that formally describes its required
behaviour, and three additional modules describing the control circuits'
behaviour, the overall design of the production cell, and some necessary
mathematical and physical background knowledge. One should note that
none of these specification modules is related to a part of the
implementation in the sense that the latter is obtained by a series of
refinements of the former. Instead, each specification module describes
a different aspect of the modelled reality, not of the implementation.

The adopted approach also discovers the senselessness of a
\QUOT{production}
cell whose purpose solely consists in circulating metal blanks, since it
is not possible to provide a goal formula that would not be also
satisfied by an empty cell. Therefore, we had to assign the travelling
crane an ability to
\QUOT{consume}
metal blanks, that is, to retransform them into unforged ones, and to
pose two separate specification goals: one for the consumer, the
travelling crane, and one for the producer, the rest of the production
cell, the latter saying in a formal notion
\QUOT{If an unforged metal blank lies on the feed belt, it will
eventually appear forged on the deposit belt}.

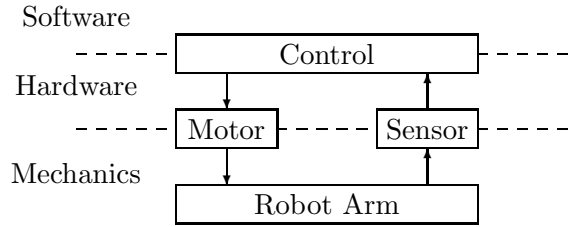
\begin{figure}
	\COMM{7.5cm x 3.0cm File f4.ps}
	{\normalsize%
%
%
%
%
%
%
%
%

\begin{center}

\begin{picture}(7.5,3.0)
\put(2.133,0.000){\framebox(4.000,0.500){{Robot Arm}}}
\put(2.133,1.000){\framebox(1.333,0.500){{Motor}}}
\put(4.800,1.000){\framebox(1.333,0.500){{Sensor}}}
\put(2.133,2.000){\framebox(4.000,0.500){{Control}}}
\put(2.800,2.000){\vector(0,-1){0.500}}
\put(2.800,1.000){\vector(0,-1){0.500}}
\put(5.467,0.500){\vector(0,1){0.500}}
\put(5.467,1.500){\vector(0,1){0.500}}
\put(0.800,2.749){\makebox(0.000,0.000){{Software}}}
\put(0.800,1.833){\makebox(0.000,0.000){{Hardware}}}
\put(0.800,0.667){\makebox(0.000,0.000){{Mechanics}}}
\multiput(0.800,1.250)(0.267,0.000){5}{\line(1,0){0.133}}
\multiput(3.467,1.250)(0.267,0.000){5}{\line(1,0){0.133}}
\multiput(6.133,1.250)(0.267,0.000){5}{\line(1,0){0.133}}
\multiput(0.800,2.250)(0.267,0.000){5}{\line(1,0){0.133}}
\multiput(6.133,2.250)(0.267,0.000){5}{\line(1,0){0.133}}
\end{picture}

\end{center}

%
%
%
%
%
%
%
%
%
%
%
%
%
\par}
	\vspace{0.5cm}

	\caption{Closed control loop including mechanical feedback}
	\label{@F4@}
	\end{figure}

The approach of predicate logic with explicit time as specification
language allows for inclusion of given technical/physical frame
requirements and thus for the treatment of systems with control loops
partly outside the hardware/software area. For example, the robot
control in some situation starts its motor to extend an arm until it
reaches a certain length, cf. figure 
\ref{@F4@}.
A verification of the subgoal that the robot arm will in fact reach the
desired length and then stop is impossible without considering the
mechanical properties of the arm involved. The same holds for the whole
production cell: to verify the ultimate specification goal that it will
produce forged metal blanks from unforged ones requires the formal
consideration of its mechanical behaviour in the proof; it does not
suffice to restrict the proof to the software, resp. hardware, aspects.

It is also possible to derive necessary requirements concerning issues
outside the hardware/software area. For example, it has been derived
that the angle between the deposit belt's starting point, the robot's
turning center, and the press has to be 90 degrees in order to deposit
the forged blanks in the right alignment angle. Thus, the deductive
approach can be extended to serve as a framework for the engineering of
the whole production cell including mechanical aspects. In a future
scenario, a mechanical engineer could be provided from his customer with
a formal requirements description of a production cell, and from the
manufacturers of the cell's machines with their formal behaviour
description. He could then develop a verified overall configuration of
the cell including its control, using the deductive approach to
integrate classical mechanical engineering tasks and software
engineering.

Finally, there was a rather surprising experience concerning the time
modelling, showing how much care is needed in formalizing the background
knowledge for the requirements specification. Consider again the control
loop of figure 
\ref{@F4@}.
It is necessary at some point of proof to show that at some time
\FORM{t_2}
the robot arm will be extended to a given length provided its length at
time
\FORM{t_1}
has been smaller. Assume that from the behavioural requirements
description of the robot we know that the arm will eventually extend to
any given length (within its limits) if its extension motor is running
long enough.

What is needed for the correctness proof of the feedback arrangement
above is, however, that there is a
\EMPH{minimal}
time in which the desired length is reached, in order to stop the
extension motor just at that point. Thus, it is not sufficient to have
rational numbers as time domain since they are not closed wrt. infima.
In fact, if the desired length the arm is to be extended to happens to
be such that it is reached if
\FORM{(t_2-t_1)^2 = 2},
then at each
\FORM{t_2 > t_1+\sqrt{2}}
the length has been reached, but there is no minimal (rational)
\FORM{t_2}.
The problem has been circumvented by including the existence of minimal
times into the requirements specification, cf. axiom 11 in figure 
\ref{@F3@}.

\SECT{Synthesis}
\COMM{how does one come from a spec to a controller?}

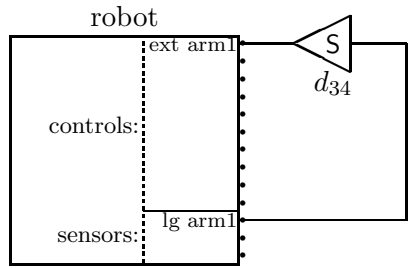
\begin{figure}
	\COMM{5.5cm x 3.5cm File f6.ps}
	{\normalsize%
%
%
%
%
%
%
%
%
\begin{center}

\begin{picture}(5.25,3.5)
\put(1.500,3.150){\makebox(0.000,0.000)[b]{robot}}
\put(1.700,0.344){\makebox(0.000,0.000)[r]{{\footnotesize sensors:}}}
\put(1.700,1.844){\makebox(0.000,0.000)[r]{{\footnotesize controls:}}}
\put(3.000,2.920){\makebox(0.000,0.000)[r]{{\scriptsize ext arm1}}}
\put(3.000,0.550){\makebox(0.000,0.000)[r]{{\scriptsize lg arm1}}}
\multiput(1.75,0)(0,.1){30}{\line(0,1){.05}}
\put(3.000,0.687){\line(-1,0){1.250}}
	\put(3.000,0.570){\line(1,0){2.250}}
	\put(5.250,0.570){\line(0,1){2.350}}
	\put(5.250,2.920){\line(-1,0){0.750}}
	\put(3.750,2.920){\line(-1,0){0.750}}
\put(4.275,2.920){\makebox(0.000,0.000){{\small\sf S}}}
\put(4.275,2.400){\makebox(0.000,0.000){$d_{34}$}}
\multiput(3.075,0.100)(0.000,0.235){13}{\circle*{0.075}}
\thicklines
\put(0.000,0.000){\framebox(3.000,3.000){}}
\put(4.500,2.545){\line(0,1){0.750}}
\put(4.500,2.545){\line(-2,1){0.750}}
\put(4.500,3.295){\line(-2,-1){0.750}}

\end{picture}

\end{center}

%
%
%
%
%
%
%
%
%
%
%
%
%
\par}
	\vspace{0.5cm}

	\caption{Control circuitery from figure @F5@}
	\label{@F6@}
	\end{figure}

Two approaches have been made to synthesize a control circuitery for the
production cell. The first approach used only the level of first order
predicates, starting from the specification as described above, and
proving its satisfiability. Figure 
\ref{@F5@}
shows an example proof of a very simple control circuitery, figure
\ref{@F6@}
shows the circuitery.

	\begin{figure}
	\COMM{16.0cm x 14.3cm File f5.ps}
	{\normalsize%
%
%
%
%
%
%
%
%
\renewcommand{\eqd}[1]{{\bf #1}}
\renewcommand{\eqr}[1]{{\bf #1}}

\renewcommand{\labelitemi}{}


Find a control circuitery to extend the robot's first arm to a given
length $d_{34}$.

Conjecture:	\\
$\exists r_0: \; \forall t_0: \; \exists t\;\;$
\begin{tabular}[t]{@{}l@{$\;$}l@{}}
& $dist\_xy(d_4,pos_1(r_0,t_0)) \.\leq d_{34}$	\\
$\ra$ & $dist\_xy(d_4,pos_1(r_0,t)) \.= d_{34}$	\\
\end{tabular}
\\
where $d_{34}=dist\_xy(d_4,d_3)$
\vspace{0.3cm}

Proof (skolem functions indicated by ``$^{\$}$''):
\vspace{0.2cm}

\begin{tabular}{@{}ll@{}}
\eqr{assumption}:
&
$dist\_xy(d_4,pos_1(r_0,t_0^{\$})) \.\leq d_{34}$	\\
\eqr{goal}:
&
$dist\_xy(d_4,pos_1(r_0,t)) \.= d_{34}$	\\
\mca{2}{\eqd{51} = \eqr{11} res \eqr{assumption},\eqr{21},\eqr{22}:} \\
&
\begin{tabular}[t]{@{}r@{$\;$}l@{}l@{$\;$}l@{}}
& $($ && $(t_0^{\$} \.\leq t_1 \.< t_2^{\$} 
	\;\ra\; extends_1(r_0,t_1))$ \\
	&& $\ra$ & $dist\_xy(d_4,pos_1(r_0,t_2^{\$})) \.= d_{34})$ \\
$\wedge$ &&& $t_0^{\$} \.\leq t_3 \.< t_2^{\$}
	\;\ra\; dist\_xy(d_4,pos_1(r_0,t_3)) \.< d_{34}$	\\
\end{tabular}
\\

\eqd{52} = split \eqr{51}:	\\
&
\begin{tabular}[t]{@{}r@{$\;$}l@{}}
& $(t_0^{\$} \.\leq t_1 \.< t_2^{\$} \;\ra\; extends_1(r_0,t_1))$ \\
$\ra$ & $dist\_xy(d_4,pos_1(r_0,t_2^{\$})) \.= d_{34}$	\\
\end{tabular}
\\

\eqd{53} = split \eqr{51}:	\\
&
$t_0^{\$} \.\leq t_3 \.< t_2^{\$}
	\;\ra\; dist\_xy(d_4,pos_1(r_0,t_3)) \.< d_{34}$
\\

\eqd{54} = \eqr{52} res \eqr{13}:	\\
&
\begin{tabular}[t]{@{}r@{$\;$}l@{}}
& $(t_0^{\$} \.\leq t_1 \.< t_2^{\$} \;\ra\; val(c_1,t_1) \.= 1)$ \\
$\ra$ & $dist\_xy(d_4,pos_1(r_0,t_2^{\$})) \.= d_{34}$	\\
\end{tabular}
\\

\eqd{55} = \eqr{54} res \eqr{31}:	\\
&
\begin{tabular}[t]{@{}r@{$\;$}l@{}}
& $(t_0^{\$} \.\leq t_1 \.< t_2^{\$} \;\ra\; val(c,t_1) \.< d_{34})$ \\
$\ra$ & $dist\_xy(d_4,pos_1(r_0,t_2^{\$})) \.= d_{34}$	\\
\end{tabular}
\\

& where 
$r_0=r(trigger(c,d_{34}),c_2,c_3,\ldots,c_8,s_1,s_2,s_3)$
\\

\eqd{56} = \eqr{55} rep \eqr{13}:	\\
&
\begin{tabular}[t]{@{}r@{$\;\;$}l@{}}
& $(t_0^{\$} \.\leq t_1 \.< t_2^{\$}
	\;\ra\; dist\_xy(d_4,pos_1(r_0,t_1)) \.< d_{34})$	\\
$\ra$ & $dist\_xy(d_4,pos_1(r_0,t_2^{\$})) \.= d_{34}$ \\
\end{tabular}
\\

& where
$r_0=r(trigger(s_1,d_{34}),c_2,c_3,\ldots,c_8,s_1,s_2,s_3)$
\\

\eqd{57} = \eqr{56} res \eqr{53}:	\\
&
$dist\_xy(d_4,pos_1(r_0,t_2^{\$}))=d_{34}$
\\

& where
$r_0=r(trigger(s_1,d_{34}),c_2,c_3,\ldots,c_8,s_1,s_2,s_3)$
\\

\end{tabular}

%
%
%
%
%
%
%
%
%
%
%
%
%

\par}
	\vspace{0.5cm}

	\caption{Example proof of a simple control circuitery}
	\label{@F5@}
	\end{figure}

\begin{figure}
	\COMM{11.5cm x 3.5 File f7.ps}
	{\normalsize%
%
%
%
%
%

\begin{center}

\begin{tabular}[b]{@{}l@{\hspace*{0.5cm}}}
length of robot arm 1:	\\
(press in upper position)	\\
\\
\\
\\
\\
\end{tabular}
\begin{picture}(6.4,3.5)

	\put(1.565,2.078){\line(2,1){0.360}}
	\put(1.925,2.258){\line(4,1){0.750}}
	\put(2.675,2.445){\line(6,1){0.750}}
\put(3.425,2.569){\line(6,1){0.700}}
\put(4.175,2.680){\circle{0.105}}
\put(4.175,3.280){\circle*{0.105}}
\put(4.225,3.270){\line(1,0){0.600}}

\multiput(1.025,2.925)(0.1,0){42}{\line(1,0){0.050}}

\put(1.575,0.550){\line(1,0){2.555}}
\put(4.220,0.160){\line(1,0){0.600}}
\put(4.175,0.565){\circle{0.105}}
\put(4.175,0.175){\circle*{0.105}}
\put(1.425,0.362){\makebox(0.000,0.000)[r]{$extends$}}
\put(5.350,0.560){\makebox(0.000,0.000)[l]{$true$}}
\put(5.350,0.165){\makebox(0.000,0.000)[l]{$false$}}
\thicklines
\put(0.800,1.825){\vector(1,0){4.400}}
\put(1.175,1.450){\vector(0,1){2.000}}
\put(5.350,1.825){\makebox(0.000,0.000)[l]{$t$}}
\put(1.025,3.000){\makebox(0.000,0.000)[r]{$d_s$}}
\end{picture}

\end{center}

%
%
%
%
%
%
%
%
%
%
\par}
	\vspace{0.5cm}

	\caption{Neclecting a safety requirement by incontinuous motion}
        \label{@F7@}
	\end{figure}
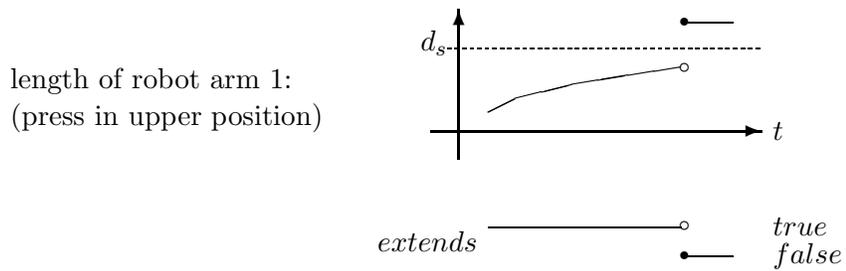

One main difficulty in finding a proof was to make explicit the
necessary assumptions about continuity of certain functions involved in
simple feedback loops. They were
\QUOT{forgotten}
in first versions of the specification and were not recognized before
the analysis of failed proof attempts. Consider, for example, the safety
requirement that the first robot arm may enter the press only if the
latter is in its middle position. Assume the control circuiteries will
stop the robot arm if approaching the press to a certain distance
\FORM{d_s}
when it is not in middle position, and prevent the press from moving off
the middle position as long as the arm remains within the distance
\FORM{d_s}.
The proof that this control meets the safety requirement, however, has
to be based on the intermediate value theorem from calculus. Figure 
\ref{@F7@}
provides a counter example if the arm movement was not continuous,
assuming the press in upper position. We had to add one instance of the
intermediate value theorem for each function required to be continuous.

The control circuitery was not really
\QUOT{synthesized}
in the sense that an actual intermediate proof goal would provide many
hints which program resp. circuitery constructs to insert. Instead, a
previously constructed circuitery was in fact verified. Moreover, to
reuse earlier parts of the proof is much easier if proofs are conducted
bottom-up, while true synthesis would require top-down (backward)
proofs. For this reason, large parts of the proof have been conducted in
a bottom-up manner, like e.g. in figure 
\ref{@F5@}.

The second approach used the experience gained during the first one to
identify higher level concepts which turned out to be valuable in
lifting specification and proof to a higher level of expressiveness. Two
new ternary logical operators were defined in terms of a restricted
second order predicate logic, see figure 
\ref{@F8@}.

	\begin{figure}
	\COMM{13.0cm x 1.6 File f8.ps}
	{\normalsize

%
%
%
%
%
%

\begin{center}

\begin{tabular}[t]{@{}l@{$\;$}l@{$\;$}l@{$\;$}l@{}}
$unt(t_0,P,Q)$ & $:\lra$
	& $\forall t_1: \;\; t_1 \.< t_0$
	& $\;\vee\; (\exists t:\;\; t_0 \.\leq t \.\leq t_1 \wedge Q(t))
		\;\vee\; P(t_1)$	\\
$ldt(t_0,P,Q)$ & $:\lra$
	& $\exists t_1: \;\; t_0 \.\leq t_1$
	& $\;\wedge\; (\forall t: \;\; 
		(t_0 \.\leq t \.\leq t_1 \ra P(t)) \;\ra\; Q(t_1))$ \\
	&&& $\;\wedge\; (\forall t: \;\; 
		t_0 \.\leq t \.\leq t_1 \ra \neg Q(t))$	\\
\end{tabular}

\end{center}

%
%
%
%
%
%
%
%
%
%

\par}
	\vspace{0.5cm}

	\caption{Application specific logical operators}
	\label{@F8@}
	\end{figure}

The concepts are borrowed from Mishra/Chandy's language
\EMPH{Unity}
\CITE{Chandy88}.
\FORM{unt(t_0,P,Q)}
means that from time
\FORM{t_0},
the unary predicate
\FORM{Q}
holds until the unary predicate
\FORM{P}
becomes true, or for ever
(\QUOT{\FORM{P} until \FORM{Q}}).
\FORM{ldt(t_0,P,Q)}
means that from time
\FORM{t_0},
if
\FORM{P}
holds long enough, then
\FORM{Q}
will eventually become true at a minimal time
\FORM{t_1}
(\QUOT{\FORM{P} leads to \FORM{Q}}).

A background theory of useful axioms about
\FORM{unt}
and
\FORM{ldt}
has been proved, including the monotonicity of
\FORM{unt}
in the second and third and of
\FORM{ldt}
in the third argument, and the anti-monotonicity of
\FORM{ldt}
in the second argument, which enabled us to include both operators into
the polarity-based non-clausal resolution rule.

Since
\FORM{unt}
and
\FORM{ldt}
reflect frequent patterns of the specification and the proof, both can
be made shorter and easier to understand by using these operators.
Figure 
\ref{@F9@}
shows the analogon to the proof of figure 
\ref{@F5@}
using
\FORM{ldt}.
One fact (61) from the background theory about
\FORM{unt}
and
\FORM{ldt}
is used.

	\begin{figure}
	\COMM{16.0cm x 7.4 File f9.ps}
	{\normalsize
%
%
%
%
%
%
%
%



\renewcommand{\eqr}[1]{{\bf #1}}

\renewcommand{\eqd}[1]{{\bf #1}}


\begin{tabular}{@{}ll@{}}

\eqd{11'}:
&
\begin{tabular}[t]{@{}r@{$\;\;$}l@{$\;$}l@{}}
$\forall r,x,t,d:$
	& \mca{2}{$robot(r,x)$}	\\
	$\wedge$ & \mca{2}{$dist\_xy(x,pos_1(r,t)) \leq d
		\leq maxlg_1$} \\
$\ra$ & $ldt(t,$ & $\lambda t_1 \!: extends_1(r,t_1),$	\\
	&& $\lambda t_2 \!: dist\_xy(x,pos_1(r,t_2)) \.\geq d)$ \\
\end{tabular}	\\

\eqd{61}:
&
$ldt(t_0,\neg P,P) \;\ra\; \exists t \;\; t_0 \.\leq t \wedge P(t)$ \\
\mca{2}{\eqd{71} = \eqr{11'} res \eqr{assumption},\eqr{21},\eqr{22}:} \\
&
\begin{tabular}[t]{@{}l@{$\;\;$}l@{}}
$ldt(t_0,$ & $\lambda t_1 \!: extends_1(r,t_1),$	\\
	& $\lambda t_2 \!: dist\_xy(x,pos_1(r,t_2)) \.\geq d_{34})$ \\
\end{tabular}	\\

\eqd{72} = \eqr{13} res \eqr{31}:	\\
& $dist\_xy(x,pos_1(r_0,t)) \.< d_{34} \;\ra\; extends_1(r_0,t_1)$ \\
& where
$r_0=r(trigger(s_1,d_{34}),c_2,c_3,...,s_3)$
as before
\\

\eqd{73} = \eqr{71} res \eqr{72}:	\\
&
\begin{tabular}[t]{@{}l@{$\;\;$}l@{}}
$ldt(t_0,$ & $\lambda t_1 \!: dist\_xy(x,pos_1(r_0,t_1)) \.< d_{34},$ \\
	& $\lambda t_2 \!: dist\_xy(x,pos_1(r,t_2)) \.\geq d_{34})$ \\
\end{tabular}	\\

\eqd{74} = \eqr{73} res \eqr{61}:	\\
& $\exists t_2: \;\; dist\_xy(x,pos_1(r,t_2)) \.\geq d_{34}$	\\

\end{tabular}


\par}
	\vspace{0.5cm}

	\caption{Proof analogon of figure @F5@, using application
			specific logical operators}
        \label{@F9@}
	\end{figure}

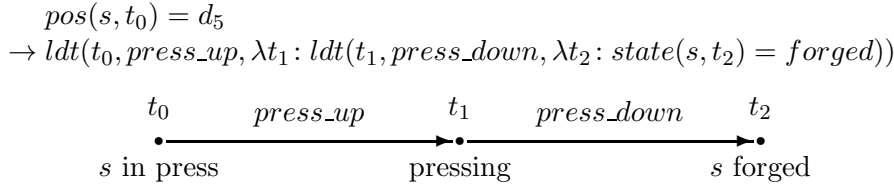
\begin{figure}
	\COMM{15.3cm x 2.1 File f10.ps}
	{\normalsize%
%
%
%
%
%
%
%
%

\begin{center}

\begin{tabular}{@{}l@{$\;$}l@{}}
& $pos(s,t_0) = d_5$	\\
$\ra$ & $ldt(t_0,press\_up,\lambda t_1\!:
	ldt(t_1,press\_down,\lambda t_2\!:
	state(s,t_2) = forged))$	\\
\end{tabular}

\begin{picture}(9.6,1.5)
\thicklines
\put(0.900,0.600){\circle*{0.100}}
	\put(0.900,0.900){\makebox(0.000,0.000)[b]{$t_0$}}
	\put(0.900,0.400){\makebox(0.000,0.000)[t]{$s$ in press}}
\put(1.000,0.600){\vector(1,0){3.800}}
	\put(2.900,0.800){\makebox(0.000,0.000)[b]{$press\_up$}}
\put(4.900,0.600){\circle*{0.100}}
	\put(4.900,0.900){\makebox(0.000,0.000)[b]{$t_1$}}
	\put(4.900,0.400){\makebox(0.000,0.000)[t]{pressing}}
\put(5.000,0.600){\vector(1,0){3.800}}
	\put(6.900,0.800){\makebox(0.000,0.000)[b]{$press\_down$}}
\put(8.900,0.600){\circle*{0.100}}
	\put(8.900,0.900){\makebox(0.000,0.000)[b]{$t_2$}}
	\put(8.900,0.400){\makebox(0.000,0.000)[t]{$s$ forged}}
\end{picture}

\end{center}

%
%
%
%
%
%
%
%
%
%
%
%
%
\par}
	\vspace{0.5cm}

        \label{@F10@}
	\caption{Modelling state transitions by
			chains of \FORM{ldt}}
	\end{figure}

Note that chains of
\FORM{ldt}s
can simulate state transitions like in the finite automaton paradigm;
figure 
\ref{@F10@}
shows an example.

L.\ Holenderski has modelled the production cell
as a kind of finite
automaton written in
\EMPH{Lustre},
allowing fully automated verification of the main requirements by state
exploration using binary decision diagrams. However, in that approach
verification cannot deal with issues that are not formalizable as
automaton properties. It would be interesting to investigate whether the
\FORM{unt}/\FORM{ldt}
approach can achieve a vertical decomposition of the model in the sense
that on the higher level only automaton properties need to be dealt with
while on the lower level the remaining properties are covered.

\SECT{Evaluation}

\SSCT{Provable properties}
\COMM{which properties are proved/can in principle be proved?}

The adopted approach makes it easy to formulate and prove all desired
liveness and safety properties. The liveness property says that each
unforged metal blank entered into the production cell will eventually
leave it forged, it has been discussed in section
\ref{Modelling the production cell}.
The safety requirements are comprised in the additional goal
\QUOT{Never any damage occurs},
where a necessary condition for any damage is provided by enumerating
all critical combinations of machines (e.g. robot/press).

A disadvantage of this approach consists in the risk of overlooking
certain possible conflict situations when writing the specification. For
example, in the
\COMM{first version of the}
informal safety requirements 
it was not required that
the feed belt may transport metal blanks only if the elevating rotary
table is empty.

Each informal safety requirement 
is a consequence of
one of the following principles:
\BI
\ITEM the avoidance of machine collisions (1, 2, 5, 6),
\ITEM the limitations of machine mobility (3, 4, 5),
\ITEM the demand to keep metal blanks from falling from great height
	(7, 9), or
\ITEM the necessity to keep the metal blanks sufficiently separate (8).
\EI
It is principally possible to base the productions cell's safety
requirements on these four principles. However, formalising the first
principle needs a complete description of the machine shapes and motion
tracks, and, moreover, a proof for each of
\FORM{n \cdot (n-1)}
pairs of machines that they will not collide, no matter how far they in
fact are separated. Since both is very expensive, we have choosen to
state the possible collision situations explicitly.

\SSCT{Explicit assumptions}
\COMM{what about assumptions about the behaviour of the
	  cell? are they explicit? are they implicit?}

Assumptions about the behaviour of a single machine as well as about the
overall configuration of the production cell are explicitly stated in
the corresponding specification module. Moreover, it is possible to
derive additional requirements on behaviour resp. configuration during
the proof, cf. section 
\ref{Modelling the production cell}.

\SSCT{Statistics}
\COMM{Messungen: Laenge der Spezifikation, Aufwand fuer Spec/Ver,
	Wiederverwendung von Bibliotheken?, usw.}

The specification comprises 8 modules with total 80 axioms, see figure
\ref{@F11@},
however, not all axioms are actually used.

	\begin{figure}
	\COMM{7.0cm x 5.2 File f11.ps}
	{\normalsize%
%
%
%
%
%
%
%
%

\begin{center}

\begin{tabular}{@{}lrl@{}}
Module & \mca{2}{No.\ of Axioms}	\\
\hline
Press & 9	\\
Robot & 24	\\
Elevating rotary table & 12	\\
Belt & 5	\\
Overall configuration & 9	\\
	incl.\ travelling crane	\\
Mathematics & 14	\\
Circuits & 8	\\
\hline
Total & 80	\\
\end{tabular}

\end{center}


\par}
	\vspace{0.5cm}

        \label{@F11@}
	\caption{Length of specification modules}
	\end{figure}

It is difficult to estimate the effort for the proof, since in parallel
to its conduction the support tool
\QUOT{Sysyfos}
had to be improved in order to be able to cope with the proof at all. As
a result of the engagement in the case study, a semi-graphic user
interface and a proof replay mechanism have been built into the support
tool; in the later phase, the restricted higher order unification for
\FORM{unt}
and
\FORM{ldt}
required some implementation work. With this relativization, the effort
e.g. for finding resp. verifying the sub-circuitery to move a metal
blank from the elevating rotary table into the press can be stated as
about 1-2 man weeks. The proof includes 210 steps without any use of
\FORM{unt}
and
\FORM{ldt}
and was the first subproof of the case study. A later proof of a
comparable task is shortened to the order of magnitude of about 1-2 man
days, due to the experience gained, especially concerning the continuity
issues discussed above.

\SSCT{Maintenance}
\COMM{is it easy to change the controller for a similar
	production cell?}

The main effort when developping a control circuitery for a different,
but similar production cell is the conduction of a new proof. It should
be easy to obtain the new formal specification, building up on the
formal terminology provided. Using the pure first order predicate logic
approach, only few parts of the original proof may be reused, depending
on the degree of similarity between both tasks. However, using the
\FORM{unt}/\FORM{ldt}
approach, a large amount of proof effort is dedicated to the
schematisation of controlling principles as background theorems which
need not be proved again, cf. e.g. theorem 61 in figure 
\ref{@F9@}
which is the heart of the proof there. It is expected that the remaining
proof effort to 
\QUOT{instantiate}
the background theorems taylored to the new cell configuration is rather
small. In any case, the necessary effort to obtain a new verified
control circuitery is still much greater than to reconfigure an object
oriented controller program, say.

\SSCT{Efficiency}
\COMM{how efficient is the controller (througput, number of
	blanks handled synchronously)?}

The paradigm of deductive program synthesis does not make any statements
about efficiency of the constructed programs. In the setting of the
production cell, moreover, efficiency does not mean short software
reaction times, but a high overall throughput rate. Following the
approach of extending software engineering methods to include also
mechanical engineering, one could estimate the
\QUOT{algorithmic complexity}
of the whole production cell. This would need the generalisation of a
complexity calculus for reactive systems. Since no recursion is
involved, the maximal work time of a metal blank could be calculated
exactly. However, a proof that the specific configuration and control of
the cell guarantee maximal thoughput seem to be as difficult as
complexity lower bound proofs for algorithmic problems.

\SSCT{Mechanical requirements}
\COMM{which conclusion about the design of the production
	cell itself are possible? (e.g.: without a sensor at
	the end of the feed belt no properties can be
	guaranteed)}

As mentioned in section 
\ref{Modelling the production cell}, 
during the synthesis proof a couple of additional requirements to the
configuration of the production cell were deduced. They mostly state
that the limitations of machine mobility allow to reach the necessary
points, e.g. that the elevating rotary table can be reached by the
robot's first arm, cf. axiom 22 in figure 
\ref{@F3@}.
Another group of requirements concerns the fitting of dimensions and
angles, e.g. that the upper position of the elevating rotary table, the
robot's first arm, and the middle position of the press must be all at
the same height.

Some conditions need not really be required but their validity would
lead to a simpler control circuitery, e.g. if it is known that the
distance from the robot's turning center to the elevating rotary table's
is greater than to the press, it is sufficient to contract the first arm
during its way to the press, otherwise the circuitery had to be prepared
for both retracting
\EMPH{and}
extending.

When operating the production cell in an
\QUOT{open}
mode, i.e. without the travelling crane, additional requirements on the
loading resp. unloading behaviour arise, e.g. the feed belt may be
loaded only if there is suffient free space available at its start. The
latter condition makes the existence of an additional feed belt sensor
necessary, either at its start or (leading to an easier and more robust
control) at its end.

Our modelling is based on the idealizing assumption that there are no
imprecisions in geometrical sizes. In practice, however, this won't be
the case, e.g. the feed belt will not deliver each metal blank exactly
to the elevating rotary table's turning center, $d3$. A model of the
production cell that takes this fact into account would have to deal
with admissable tolerance intervals, stating e.g. that the robot's first
arm will safely grab the metal blank if it lies within the area
\FORM{d_3+x}
with
\FORM{\mid\mid \.x \mid\mid < \varepsilon_3}.
Each machine may add its own inaccuracy to the tolerance interval, but
may also decrease the interval in some respect due to some alignment
effect, e.g. at photoelectric cells. Then, one has to require additional
that the tolerance intervals are small enough to allow proper operation.
E.g. the tolerance interval of a metal blank's position in the press
contains the sum of tolerances of the robot's first arm, the elevating
rotary table, the feed belt, and the (external) feed belt loading
device; it must be ensured that this deviation is small enough to allow
safe pressing of the blank.

\SECT{Conclusion}

Our experiences with the production cell case study seem to confirm the
following theses:
\BI
\ITEM \EMPH{A good requirements specification should consist of a
        collection of almost obvious facts in formal notation.}
        The absense of need for executability provides the freedom to
        state formal requirements as an almost direct translation of
        natural language formulation. The former can be validated
        against the latter in a local manner.
\ITEM \EMPH{Requirement specification modules describe different aspects
        of the modelled reality, not of the implementation.}
        In contrast to design specification modules, the former do not
        refine into implementation modules, they are rather orthogonal
        to them.
\ITEM \EMPH{Predicate logic can be seen as
	\QUOT{assembler language}
	for specifications.}
        It is desirable to build higher language constructs upon it in
        order to come to more concise specifications as well as proofs.
\ITEM \EMPH{The level of formal description can be lifted as high as
        purely technical issues are involved.}
        There seems to be no reason to stop within the level of software
        engineering, rather, the logic-based methods can serve as a
        framework for a verified overall engineering. This has been
        demonstrated by our treatment of the production cell which lies
        entirely in the technical area and whose specification included
        the topmost goal (production of forged metal blanks). On the
        other hand, if the topmost goal is non-technical, like e.g. in a
        medical information system, our approach is not fully
        applicable.
\ITEM \EMPH{There are only a couple of adequate levels of description.}
        Our experience has shown that the decision to choose a
        non-discrete time modelling necessarily implies a description
        based on continuous time and continuous functions; there seems
        to exist no intermediate level (e.g. of rational time and
        arbitrary functions). A more realistic approach could use
        differentiable functions. While in the former approach, for
        example, a motor is assumed to run with full speed immediately
        after it has been started, the latter approach allows to reason
        about accelerations and starting velocities. While not urgently
        required for the production cell case study, this level of
        description becomes unavoidable when dealing with time critical
        applications eg. from the area of vehicle control systems where
        it is vital to talk about acceleration and brake times.
\EI

\SECT{References}

\BD
\DITM{Burghardt94}
	Deduktive Synthese der Steuerung einer Fertigungszelle,
	Jochen Burghardt,
	GMD working papers,
	to appear 1994
\DITM{Chandy88}
	Parallel Program Design, A Foundation,
	Misra, J., Chandy, J.,
	Addison-Wesley,
	1988
\DITM{Manna80}
	A Deductive Approach to Program Synthesis,
	Zohar Manna, Richard Waldinger,
	in: ACM Transactions on Programming Languages and Systems,
	Vol. 2
	No. 1,
	p. 90-121,
	Jan 1980
\DITM{Manna86}
	Special Relations in Automated Deduction,
	Zohar Manna, Richard Waldinger,
	in: Journal of the ACM,
	p. 1-59,
	Jan 1986
\ED

\end{document}